\documentclass[reprint,superscriptaddress]{revtex4-1}
\usepackage[utf8]{inputenc}
\usepackage[T1]{fontenc}
\usepackage[ngerman, english]{babel}
\usepackage{graphicx}
\usepackage{xcolor}
\usepackage{multirow}
\usepackage{amsmath,amssymb}
\usepackage{float}
\usepackage{mathptmx}
\usepackage{etoolbox}
\usepackage{soul}
\makeatletter
\def\@email#1#2{%
 \endgroup
 \patchcmd{\titleblock@produce}
  {\frontmatter@RRAPformat}
  {\frontmatter@RRAPformat{\produce@RRAP{*#1\href{mailto:#2}{#2}}}\frontmatter@RRAPformat}
  {}{}
}%
\makeatother

\begin{document}
\onecolumngrid
\textcolor{blue}{\textit{This is the author’s peer reviewed, accepted manuscript. However, the online version of record will be different from this version once it has been copyedited and typeset. After it is published, it will be found at https://doi.org/10.1063/5.0207235. PLEASE CITE THIS ARTICLE AS DOI: 10.1063/5.0207235}}

\title{Negative longitudinal resistance of monolayer graphene in the quantum Hall regime}

\author{Alexey A. Kaverzin}
 \thanks{corresponding author: akaverzin@g.ecc.u-tokyo.ac.jp}
\affiliation{Department of Applied Physics, The University of Tokyo, Tokyo 113-8656, Japan}
\affiliation{Institute for AI and Beyond, The University of Tokyo, Tokyo 113-8656, Japan}
\author{Shunsuke Daimon}
\affiliation{Department of Applied Physics, The University of Tokyo, Tokyo 113-8656, Japan}
\affiliation{Institute for AI and Beyond, The University of Tokyo, Tokyo 113-8656, Japan}
\affiliation{Quantum Materials and Applications Research Center, National Institutes for Quantum Science and Technology, Tokyo 152-8552, Japan}
\author{Takashi Kikkawa}
\affiliation{Department of Applied Physics, The University of Tokyo, Tokyo 113-8656, Japan}
\author{Tomi Ohtsuki}
\affiliation{Physics Division, Sophia University, Chiyoda, Tokyo 102-8554, Japan}
\author{Eiji Saitoh}
\affiliation{Department of Applied Physics, The University of Tokyo, Tokyo 113-8656, Japan}
\affiliation{Institute for AI and Beyond, The University of Tokyo, Tokyo 113-8656, Japan}
\affiliation{WPI Advanced Institute for Materials Research, Tohoku University, Sendai 980-8577, Japan}

\begin{abstract}
In the quantum Hall regime the charge current is carried by ideal one-dimensional edge channels where the backscattering is prohibited by topology. This results in the constant potential along the edge of the Hall bar leading to zero 4-terminal longitudinal resistance $r_{xx}$. Finite scattering between the counter-propagating edge states, when the topological protection is broken, commonly results in $r_{xx} > 0$. However, a local disorder, if allowing intersection of the edge states, can result in a counter-intuitive scenario when $r_{xx}<0$. In this work we report the observation and a systematic study of such unconventional negative longitudinal resistance seen in an encapsulated monolayer graphene Hall bar device measured in the quantum Hall regime. We supplement our findings with the numerical calculations which allow us to outline the conditions necessary for the appearance of negative $r_{xx}$ and to exclude the macroscopic disorder (contamination bubble) as the main origin of it.

\end{abstract}

\maketitle

Quantum Hall regime is one of the possible ways to realize a dissipationless transport, which has already provided progress in the resistance metrology area \cite{Kruskopf2018}. Moreover, it promises a practical advantage for a lossless information transfer and can potentially be used as a platform for the realization of electron based quantum computing \cite{PRLWillet2013, Nakamura2020, Bauerle_2018, PhysRevXDuprez2019} and quantum optics \cite{Ji2003}. Yet, the current level of control of the properties of the edge channels is not sufficient for the realization of such technology and, therefore, further properties of the edge states and various possibilities of the manipulation of them remain to be studied for progress in this area.

\begin{figure}[h]
    \centering
    \includegraphics[width=0.47\textwidth]{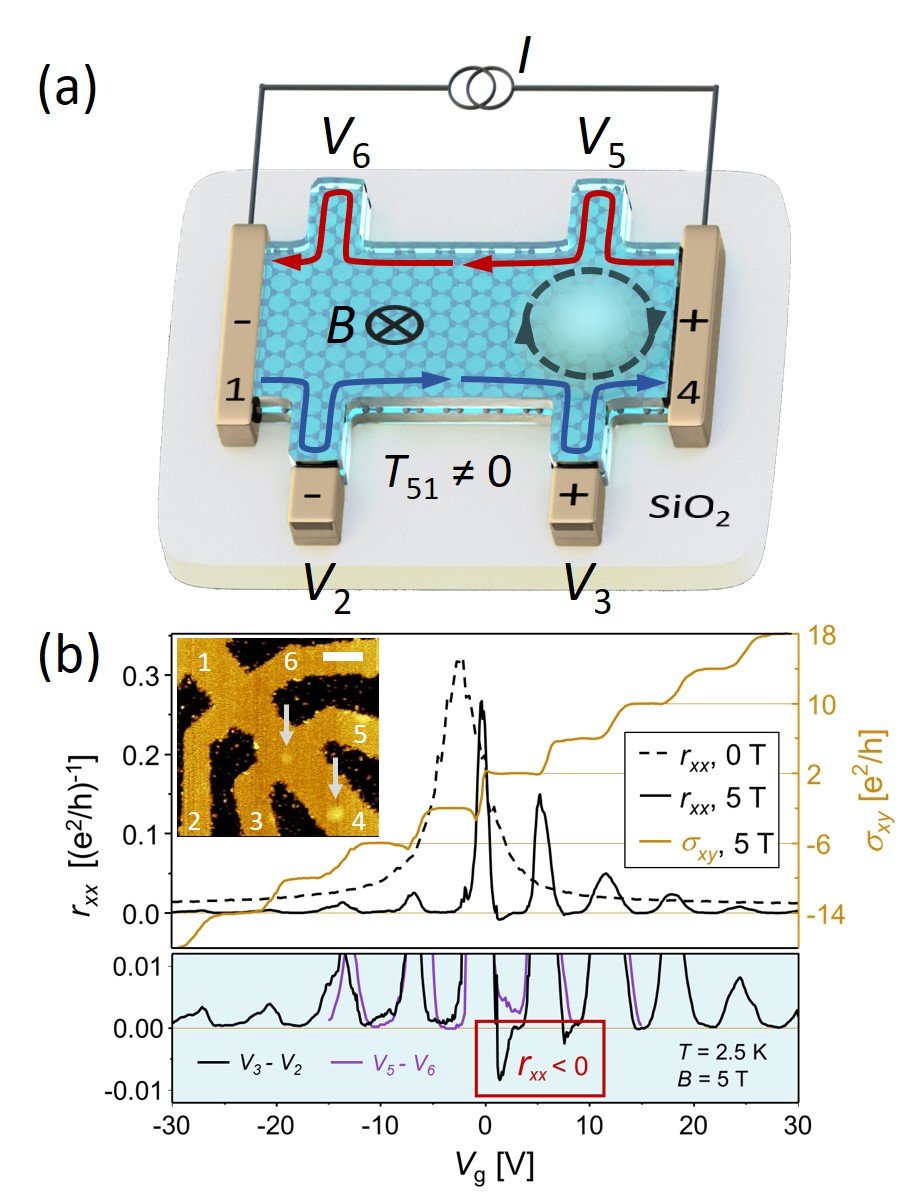}
    \caption{(a) Schematic illustration of the $h$-BN encapsulated graphene Hall bar with the edge (red and blue) and localized (grey) states which can result in a finite $T_{51}$. The arrows indicate the direction of the electron movement.
    (b) $r_{xx}=(V_3-V_2)/I$ (black) and $\sigma_{xy}$ calculated from $r_{xy}=(V_5-V_3)/I$ (pale brown) of the device measured at $2.5\,$K under $0\,$T (dashed line) and $5\,$T (solid lines) out-of-plane magnetic field $B$. Bottom sub-panel shows $(V_3-V_2)/I$ (same as in the main panel, black) and $(V_5-V_6)/I$ (purple) under applied $5\,$T with amplified scale. Inset: AFM image of the device. Contact pairs 3-2 and 5-3 are used as voltage probes for the $r_{xx}$ and $\sigma_{xy}$ measurements (AC current is supplied between 4 and 1) shown in the main panel. Arrows point to the two contamination bubbles. The scale is given by a $500\,$nm long white bar.}
    \label{fig1:my_label}
\end{figure}

In a homogeneous system the edge state, formed in the quantum Hall regime, flows strictly along the equipotential edge. This means that voltage electrodes from the same side (as contacts 3 and 2 in Fig.$\,$1(a)) couple to the same potential which guarantees $r_{xx}$ to be exactly $0$. $r_{xx}$ here is a 4-terminal longitudinal resistance defined as a voltage difference measured along the sample edge and divided by the applied current $(V_3-V_2)/I$. Commonly the coupling between the states, flowing along the opposite edges of the channel (as red and blue states in Fig.$\,$1(a)), results in a finite positive $r_{xx}$. However, when the coupling is localized and has an appropriate strength and geometry, it can lead to an exotic scenario with 4-terminal resistance becoming negative \cite{Buttiker1988}. This is a direct consequence of the potential order being reversed due to the crossing of the top and bottom edge states so that electrodes from the same side couple to different potentials. Such intersection of the edge states can take place at the localized state of an appropriate macroscopic size, comparable to the width of the channel, Fig.$\,$1(a). 

Graphene is an exceptional material, where quantum Hall effect is observed at much softer limitations on the magnetic field and temperature as compared to other semiconductor structures \cite{NovoselovScience2007,Kruskopf2018}. However, in terms of the quality of the quantization, commonly used Hall-bar-shaped graphene samples lose in comparison with high mobility semiconductors. In recent years it has been suggested that the origin of this difference lies in the non-homogeneous electrostatic screening and in the non-intentional charged impurities at the graphene edges, both of them resulting in the distinctive doping of the edges. Such non-homogeneous doping profile ultimately affects the manifestation of the quantum Hall effect since the counter-propagating edge states can appear even at the same edge of the graphene Hall bar \cite{Marguerite2019,Moreau2021,PhysRevBMoreau2021,PhysRevBMoreau2022}. If these closely positioned counter-propagating states are coupled via a single microscopic impurity, it can also result in the reversal of the potential order and manifest itself as $r_{xx}<0$ (Sec.~8 of the supplementary material).

Negative resistance in the quantum Hall regime, in fact, has been seen in various two-dimensional systems \cite{CHANG1988769,PRLTimp1989,JPCMHerfort1995,PRBTakagaki1996,SSCTakagaki1998,PhysRevBBours2017,NatPhysLi2019,PRLFu2020,NNanoShi2020,NatComBarrier2020}. In particular, in Ref.~\cite{CHANG1988769} the appearance of negative longitudinal resistance (seen in $80-200\,$nm wide AlGaAs/GaAs-based channels) was explained as the result of the quantum interference of the edge states, however, it was not theoretically elaborated and was not ascribed to any of the properties of the channel. Furthermore, there are several reports on negative bend resistance in the quantum Hall regime in AlGaAs/GaAs-based structures \cite{PRLTimp1989,JPCMHerfort1995,PRBTakagaki1996,SSCTakagaki1998}, where non-local signal is collected in a cross-shaped circuit geometry and is associated with direct scattering between the edge states from the opposite sides of the channel.

Negative longitudinal resistance $r_{xx}$ in a 4-terminal Hall-bar geometry can be spotted in the experimental reports on graphene \cite{PhysRevBBours2017} and other layered materials \cite{NNanoShi2020}, yet in these reports it is not studied and is fully excluded from the discussion.

In this work we report the experimental observation of the negative longitudinal resistance in monolayer graphene studied under varying Fermi level, magnetic field and temperature. Our observations cannot be attributed to direct scattering between the edge states from the opposite sides of the channel due to the considerable separation between the channel sides ($\simeq500\,$nm), thus, invoking an alternative explanation. We perform the tight-binding calculations where we demonstrate that the localized doping potential with appropriate dimensions can mediate the transmission between the counter-propagating edge states and result in $r_{xx}<0$. We also conclude that experimentally found contamination bubbles cannot explain our results due to their spatial position and undersized dimensions. We, therefore, suggest that the most likely origin of the negative resistance in our sample is local impurities that are unseen by AFM. Such impurities can couple the counter-propagating edge states from the same side of the channel, appearing there due to unintentional doping of the edges (see Sec.~8 in the supplementary material and Fig.~2(e)). Our conclusions highlight negative resistance both as a tool for identifying/characterizing microscopic- and macroscopic-size contamination in the channel and as an additional functionality for the manipulation of the electron transport in the quantum Hall regime.

Graphene resistance $r_{xx}$ measured as a function of the gate voltage $V_{\rm{g}}$ at zero magnetic field $B=0\,$T and at $2.5\,$K (Fig.$\,$1(b)) shows typical behavior with maximum in $r_{xx}(V_{\rm{g}})$ (Dirac point) being shifted from $0\,$V. This shift indicates the residual doping estimated to be $\approx1.5\cdot10^{12}\,$cm$^{-2}$. Carrier mobility was found to be $\mu\approx1.5\,$m$^2/$(Vs). Applying magnetic field in the out-of-plane direction causes the formation of the Landau levels and results in a quantum Hall effect. The transverse conductance of monolayer graphene becomes quantized to the values $\sigma_{xy} = 4e^2/h\cdot(i+1/2)$, where $i$ is an integer number. This is coherent with our observations, Fig.$\,$1(b), confirming that the channel is made from monolayer graphene.

We notice that the $r_{xx}$ maximum associated with zeroth Landau level (at $V_{\rm{g}}\approx-0.2\,$V) is shifted with respect to the Dirac point position found to be at $\approx-2.1\,$V at $B=0\,$T. In the quantum Hall regime measured $r_{xx}$ and $\sigma_{xy}$ represent edges of the device rather than the bulk of it \cite{Marguerite2019}, which implies that the edges of our device are hole-doped with respect to the bulk. Furthermore, the $r_{xx}$ maxima are much higher on the electrons side of the $r_{xx}(V_{\rm{g}})$ dependence as compared to the holes side, thus also suggesting inhomogeneous doping profile in the sample.

As a central result to this work, we found that at $B=5\,$T near both $V_{\rm{g}}\approx1\,$V and $V_{\rm{g}}\approx8\,$V graphene resistance becomes negative (Fig.$\,$1(b), bottom), whereas it is expected to be just zero or positive in the presence of finite scattering. Furthermore, longitudinal resistance is found to be negative only when measured with electrodes $3$ and $2$, but stays positive for the pair $5$ and $6$. Observation of $r_{xx}<0$ is intriguing and usually indicates the ballistic character of the carrier movement. There are several known phenomena that result in negative resistance such as magneto-focusing of the carriers \cite{Taychatanapat2013}, snake states \cite{Rickhaus2015}, so-called bend-resistance \cite{Beenakker1989,Timp1988} and viscous electron flow \cite{Bandurin}. However, these effects occur commonly at low or no magnetic fields and cannot explain our observations under well-developed quantum Hall regime.

The electron transport in mesoscopic conductors can be described  within the Landauer-Büttiker formalism, where from the set of transmission and reflection coefficients between the electrodes (carrier reservoirs) one can derive the values for all measurable resistances. Within this formalism the 4-terminal resistance is given by:
\begin{equation}
   R_{41,56}={h}/{e^2}\,{(T_{54}T_{61}-T_{51}T_{64})}/{D},
   \label{Eq:R4T01}
\end{equation}
where current(voltage) is supplied(measured) between electrodes $1$ and $4$
($5$ and $6$) labeled as in Fig.$\,$1(a); $T_{ij}$ is the probability for the carrier to be transmitted from the electrode $j$ to the electrode $i$ and $D$ is a subdeterminant of the matrix formed by the coefficients in the set of equations for the electric current \cite{Buttiker1988}. From Eq.~(\ref{Eq:R4T01}) it is apparent that 4-terminal resistance can in fact become negative when the probability for the carrier to get from electrode $1$($4$) to the electrode $5$($6$) is larger than the probability to get to the closer electrode $6$($5$). This is impossible in case of the diffusive transport, whereas in the ballistic or quantum Hall regimes $T_{54}T_{61}<T_{51}T_{64}$ can actually be realized, necessarily implying the crossing of the carrier path from electrode $1$ to electrode $5$ with the path from electrode $4$ to electrode $6$.

These conditions for the observation of $r_{xx}<0$ in the quantum Hall regime were formulated in Ref.~\cite{Buttiker1988}, where Büttiker proposed the localized state as an appropriate coupling between the edge states. Furthermore, he demonstrated that the magnitude of the negative resistance sharply decays with the number of the edge states. This is in fact reproduced in Fig.$\,$1(b), where the negative resistance is most pronounced for $i=0$ ($V_{\rm{g}}\approx1\,$V), much weaker for $i=1$ ($V_{\rm{g}}\approx8\,$V) and practically vanishes for $i\geq2$. Out of many available edge states the majority is unaffected by the coupling via the localized state, thus, on average any effect of the localized state is expected to decay with increasing $i$.

Since counter-propagating edge states can appear either on opposite sides or the same side of the channel, possible origins of the transmission mediating localized state in graphene Hall bar are: (1) the boundary between differently doped regions that is present around the macroscopic contamination bubble and (2) a microscopic-size/single impurity positioned near the edge. In the AFM image of our device a few bubbles are noticed and two of them are in close vicinity to the electrode $3$ (Fig.$\,$1(b), inset). As we show schematically in Fig.$\,$1(a) in the presence of such contamination bubble the bottom edge state (blue) can traverse the device, enter the top edge state and continue towards the contact $5$ which will result in $T_{51}\neq0$ and $R_{41,56}<0$. Remarkably, for the given chirality (propagation direction) of the edge states and sample geometry as in Fig.~1 we expect $R_{41,32}\geq0$ due to $T_{25}=0$. Therefore, already from the qualitative argumentation we conclude that (1) the non-equivalence between $R_{41,56}$ and $R_{41,32}$ is expected and reproduced experimentally (bottom sub-panel of Fig.$\,$1(b)) and (2) the measured $r_{xx}=R_{41,32}<0$ cannot be explained by the detected by AFM contamination bubble.

\begin{figure}[h]
    \centering
    \includegraphics[width=0.47\textwidth]{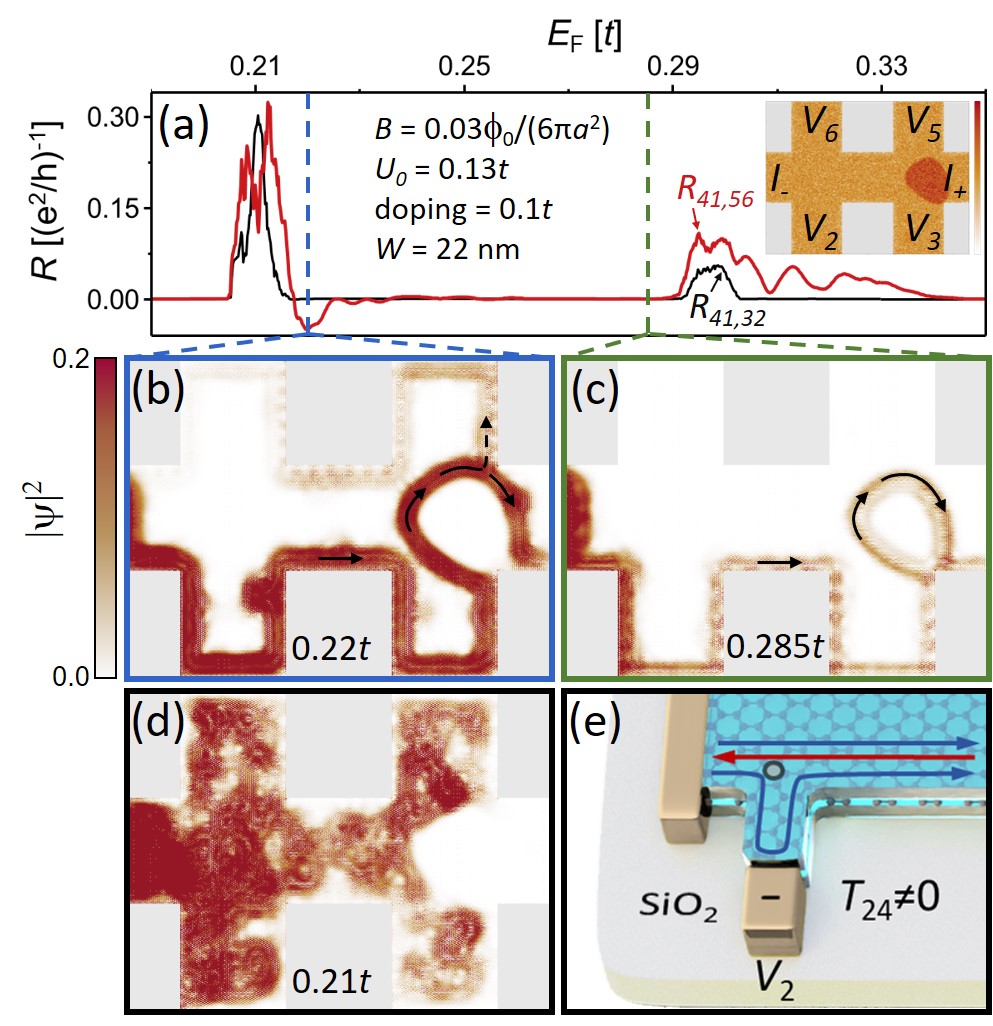}
    \caption{(a) Calculated longitudinal resistance of the Hall bar shaped graphene as a function of the Fermi energy $E_{\rm{F}}$ in units of the hopping parameter $t$. Parameters used in the calculation are: magnetic field $B=0.03\phi_0/(6\pi a^2)$, where $\phi_0$ is magnetic flux quantum and $a$ is lattice constant; strength of random potential $U_0=0.13t$ and doping strength $0.1t$. Inset gives the shape of the graphene region (channel width $W=22\,$nm) with current and voltage electrodes used for calculating resistance being labeled. The color represents the onsite potential, where darker region corresponds to a local doping due to the contamination bubble. Panels (b)-(d) represent the density of the scattering states injected from the left contact for corresponding Fermi energies. (e) Schematic representation of the edge states belonging to the same side of the channel and coupled via a local impurity.}
    \label{fig2:my_label}
\end{figure}

To unambiguously confirm and quantify the influence of the bubble on the edge states' transport we set up a tight-binding model by using the KWANT module \cite{Groth_2014} (Sec.~5 in the supplementary material). The simulated circuit and geometry of graphene sample are shown in the inset of Fig.$\,$2(a). The presence of the bubble was taken into account as the local onsite potential shown as a dark colored region. The strength of the coupling between the localized state and edge states is controlled via the set of distances between the doped region and the channel edges (Sec.~6 in the supplementary material). Together with the strength of the doping these distances are varied in order to find the conditions when negative resistance appears. Orientation of $B$, current direction and the edge state chirality are taken from the experiment.

When using the position and dimensions of the doped region to be as close as possible to the visible in AFM bubble, we found in our simulation both $R_{41,56}\geq0$ and $R_{41,32}\geq0$, thus, confirming that the bubble is unlikely to be the origin of $r_{xx}<0$. Nevertheless, we varied the size of the doped region in order to determine the conditions at which either $R_{41,56}<0$ or $R_{41,32}<0$ appear. In Fig.$\,$2(a) we plot an example of the calculations when $R_{41,56}<0$ is seen and is well-pronounced within a relatively large range of energies. To understand the underlying mechanism and the role of the doping potential we plot in panels (b)-(d) the density of all scattering states that depart from the left current contact at a given Fermi level $E_{\rm{F}}$. When the Landau level is crossed either in the main channel or in the doped region the resistance is positive and is governed by the diffusive transport. This is evidenced by the wave function amplitude (panel (d)) that is present everywhere across the channel indicating efficient scattering. Panel (c) corresponds to no coupling between the localized and top states which is topologically equivalent to the absence of the doping potential and results in $R_{41,56}=R_{41,32}=0$. The relevant case of partial coupling is shown in panel (b): the bottom edge state continues along the boundary of the doped region and near the top it is partially transmitted to the contact $V_{5}$. Since the bottom edge state has the low potential, the voltage $V_{5}$ becomes lower than $V_{6}$ and results in $R_{41,56}=(V_5-V_6)/I<0$. At the same time for a given sample geometry we find no substantial negative $R_{41,32}$. These findings validate our earlier qualitative conclusions that contamination bubble can lead to $R_{41,56}<0$, yet cannot result in $R_{41,32}<0$ and, therefore, cannot explain experimentally observed $r_{xx}<0$.

Another experimental evidence that suggests omitting contamination bubbles from consideration is the observation of $r_{xx}<0$ in an adjacent region of the device, where no bubbles are visible in the AFM scan (Sec.~3 in the supplementary material). Together with both our theoretical considerations and with the observation of the edge doping it suggests an alternative explanation for the origin of the coupling between the counter-propagating edge states. We propose here that measured $r_{xx}<0$ originates from the impurity, unseen by AFM due to small size and located in the vicinity of the contact $V_2$. Due to the non-intentional doping, counter-propagating states can appear at the same edge, Fig.~2(e), and appropriately placed impurity in between them can provide $T_{24}\neq0$ and, thus, explain $r_{xx}=R_{41,32}<0$. In this case the top side of the channel is unaffected, which is coherent with our experimental results, since $V_5-V_6$ stays positive in the whole range of $V_{\rm{g}}$, Fig.$\,$1b (also Sec.~4 in the supplementary material). Furthermore, we note that from Eq.~(\ref{Eq:R4T01}) and Figs.$\,$1(a) and 2(e) it is apparent that change of the chirality changes the condition for the observation of the $r_{xx}<0$. This explains that under fixed $B$ the effect is seen only for one type of the carriers ($E_{\rm{F}}>0$), although this is valid for both microscopic impurity and macroscopic bubble providing the coupling.

\begin{figure}[t]
    \centering
    \includegraphics[width=0.47\textwidth]{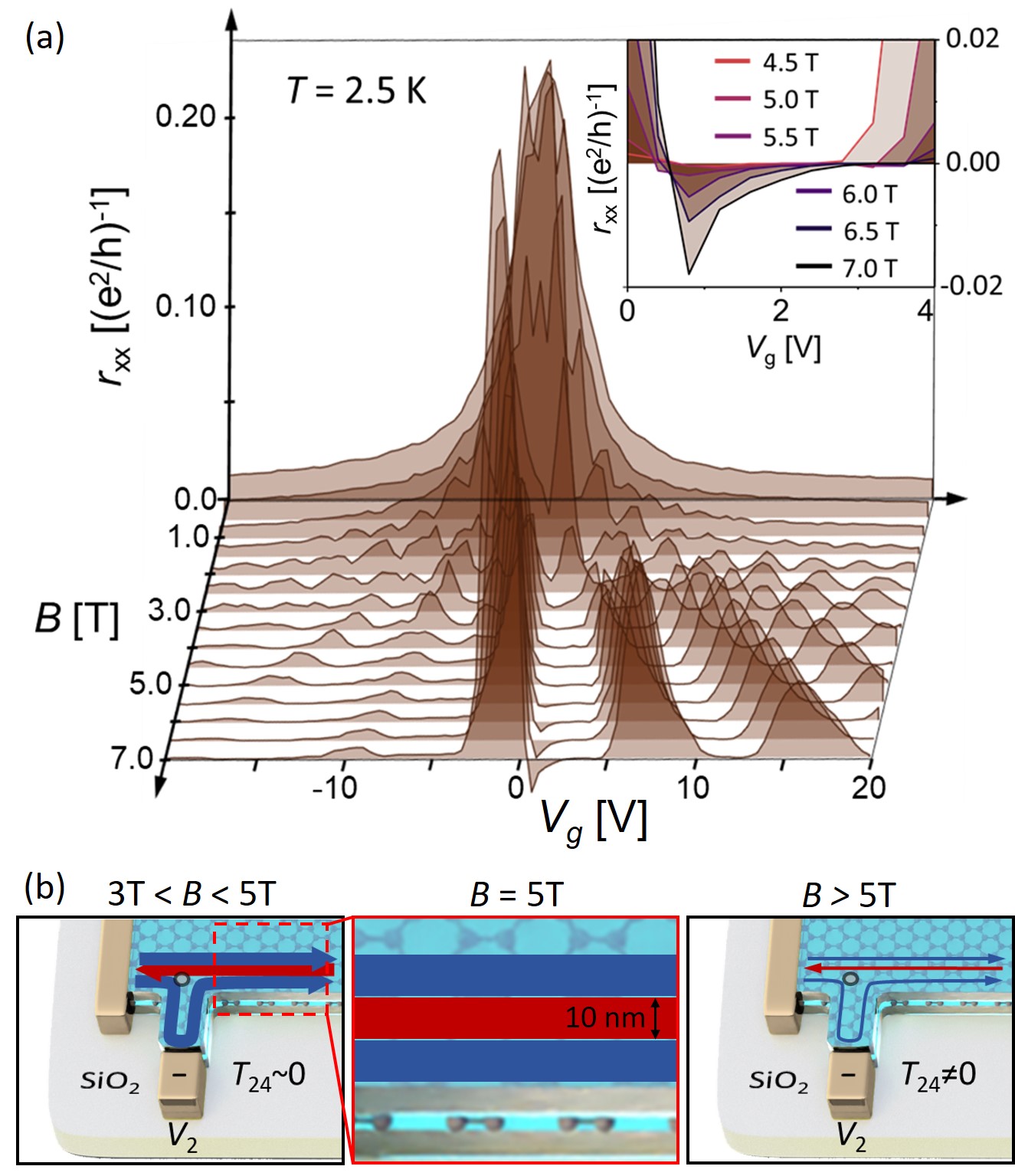}
    \caption{(a) Longitudinal resistance as a function of the applied gate voltage for different magnetic fields. The areas between each of the plotted curves and its projection to the plane $r_{xx}=0\,\Omega$ are shaded with light-brown. Inset: $r_{xx}$-$V_{\rm{g}}$ plane projection of the main panel for the fields above $4.5\,$T. (b) Schematic illustrations of the transport regimes depending on applied $B$. The width of the arrows represents the width of the edge states.}
    \label{fig3:my_label}
\end{figure}

In order to characterize further the experimental conditions under which $r_{xx}<0$ is observed we vary the out-of-plane magnetic field. In Fig.$\,$3(a) $r_{xx}(V_{\rm{g}})$ is plotted in a 3D waterfall plot for different $B$. Maxima in $r_{xx}$ spread out from the Dirac point with the increasing $B$ in accordance with the expected positions of the Landau levels. The minima in longitudinal resistance become pronounced already at $3\,$T, yet the negative resistance appears only at $B\gtrsim5\,$T. The size of the edge states is defined by the magnetic length, which in graphene at $B=5\,$T reads $l_B=\sqrt{{\hbar}/{(eB)}}\approx26\,$[nm]$/\sqrt{{B}/{(1\,[\rm{T}])}}\approx10\,$nm. It implies that the relevant edge states are separated by $\approx10\,$nm (Fig.$\,$3(b), middle). For $3\,$T$\,<B<5\,$T the neighboring counter-propagating states overlap due to their finite width  (Fig.$\,$3(b), left), which results in a diffusive non-localized scattering between them. The presence of the additional local impurity has a minor influence in this case. Increasing the field reduces the width of the edge states, removes the overlap between them, and promotes the local impurity coupling as the only source of the scattering (Fig.$\,$3(b), right). Transport regime in this case is considered to be ballistic, since the scattering appears only at the impurity.

In this work we report an observation and analysis of a non-conventional negative longitudinal resistance in $h$-BN encapsulated monolayer graphene in the quantum Hall regime. With the help of the tight-binding calculations we have shown an explicit example of sample geometry where the optimized coupling between the edge states results in the appearance of negative $r_{xx}$. We have also identified the role of chirality of the edge states, which determines the carrier type and spatial position of the coupling, necessary for the observation of $r_{xx}<0$. For the case of our sample, we excluded visible contamination bubbles as the likely candidate for the origin of the observed effect. Instead we concluded that the most probable explanation is the local microscopic impurity which couples the counter-propagating edge states belonging to the same side. The effect is only seen when the system is in a well-developed quantum Hall regime with sufficient insulation between these edge states separated by $\approx10\,$nm. 

In line with the recent literature \cite{Marguerite2019,Moreau2021} our observations offer independent evidence for the non-trivial formation of the edge states in graphene Hall bars. This finding holds broad implications for understanding of not only quantum Hall effect but also other quantized transport phenomena (i.e. quantum anomalous Hall effect \cite{Serlin2020}) in all two-dimensional materials. Furthermore, negative resistance can be used as a control parameter for targeted placement of the coupling between the neighbouring edge states, when designing the circuit for the quantum optics experiments. Finally, our analysis suggest that the appearance of the negative resistance is extremely responsive to the position and size of the associated local impurity thus promoting negative resistance as a sensitive tool for identifying the presence and/or characteristics of the edge disorder.

\section*{\begin{flushleft}Supplementary Material\end{flushleft}}
Supplementary material includes: detailed description of the sample fabrication, tight-binding model, supporting numerical and experimental results, and an extended discussion on the proposed coupling mechanism.

\section*{\begin{flushleft}ACKNOWLEDGMENTS\end{flushleft}}
We would like to acknowledge Prof. Y. Iwasa, Y. Itahashi and Y. Dong for the help in the sample fabrication. This work was supported by JST-CREST (JPMJCR20C1 and JPMJCR20T2), Grant-in-Aid for Scientific Research (JP19H05600, JP20H02599, and JP22K18686) and Grant-in-Aid for Transformative Research Areas (JP22H05114) from JSPS KAKENHI, Japan, and Institute for AI and Beyond of the University of Tokyo.

\section*{\begin{flushleft}AUTHOR DECLARATIONS\end{flushleft}}
\subsection*{\begin{flushleft}Conflict of Interest\end{flushleft}}

The authors have no conflicts to disclose.

\subsection*{\begin{flushleft}Author Contributions\end{flushleft}}

\section*{\begin{flushleft}DATA AVAILABILITY\end{flushleft}}

The data that supports the findings of this study are available from the corresponding author upon reasonable request.
\\
\\
\textit{Copyright 2024 Author(s). This article is distributed under a Creative Commons Attribution-NonCommercial-NoDerivs 4.0 International (CC BY-NC-ND) License.}

\end{document}